\documentstyle{article}
\begin{document}
{\centerline {\Large  {\bf How to test vector nature of gravity}}}

\bigskip
\hspace {4 mm} { \parbox {10 cm} {{\centerline  { Igor E. Bulyzhenkov }}

{\small

\bigskip
\noindent
{\it Institute of Spectroscopy, RAS, Troitsk,
142092 Moscow reg., Russia}

\bigskip

The covariant scheme is proposed to couple gravity and
electrodynamics in pseudo-Riemannian four-spaces with
electromagnetic connections. Novel dynamics of the charged
particle and electromagnetic dilation-compression of its proper time
can be tested in non-relativistic experiments.
The vector equations acknowledge unified photon waves
without metric modulations of flat laboratory space.

\bigskip
\noindent PACS numbers:  04.50.+h,  12.10.-g   }   }}

\bigskip
The common expectation for the forthcoming search [1] of
gravitation waves is that gravity has a tensor nature and
gravitational or metric waves ought to be quite different
from vector electromagnetic waves.
The unorthodox paradigm of curved three-space was
successfully employed in the last century to explain the
precise gravitational observations [2-3],  but divorced
electrodynamical and gravitational forces: electric charges
do not disturb 3D geometry.

Nonetheless the similarity of Newton and Coulomb interactions
may suggest a natural similarity or identity of gravitons and
photons, which are responsible for these interactions. An alternative
opportunity to explain  gravitational observations is to keep
flat three-space but to derive the pseudo-Riemannian four-interval
from a nonlinear relation,
$ds^2 = [d\tau (ds)]^2 - \delta_{ij}dx^idx^j$.
Only three-spaces $x^i$ with constant curvature, including flat space
with $\gamma_{ij} = \delta_{ij}$, are compatible with the
well-tested conservation of a system three-momentum at all
space points.

The present scheme is based on material states of different
charged objects in  their proper four-spaces $x^\mu_{_K}$ with
different pseudo-Riemannian metrics. But evolution of material
objects may be compared and observed within 3D inter\-sections
$x^i_{_K} = x^i$ of the proper four-spaces,
because all sub-spaces $x^i_{_K}$ keep universal Euclidean geometry
and may form laboratory three-space. Flat laboratory space $x^i$
and a universal  time interval $dt = \pm |dx^o|$
are commom for all  extended particle-field objects, while any
curved four-space
$x^{\mu}_{_N}$ may be associated only with one selected object N.
The accepted gravitation of the point charges in common curved four-space
differs from  the novel scheme, called vector electrogravity, where
 non-relativistic electrodynamic relations may be used to test gravitation
of the extended charges and their point sources.

In order to prove joint roots for gravitational and electrodynamic
fields one has to derive all variation equations  in a joint
vector form and test the derived dynamical equations in practice.
Let all fields from external sources be based on the unified
forming-up four-potentials $a_{_K\mu}$
and contribute jointly into the proper tetrad of any selected object N,
\begin {equation}
e^\alpha_{_N\mu}  = \delta^\alpha_{\mu}
 + \delta^{\alpha o} {\sqrt {1-\delta_{ij}v^i_{_N}v^j_{_N}}}
U^{\neq {_N}}_\mu ,
\end {equation}
where
 $U^{\neq _N}_\mu(x)
 $ = $   \sum_{_K}^{_K\neq _N}
( -G m_{_K} +  m^{-1}_{_N}q_{_N}q_{_K} )a_{_K\mu }$.
One may verify that the proper pseudo-Riemannian tensor,
$ {g}^{_N}_{\mu\nu} \equiv
\eta_{\alpha\beta}e^\alpha_{_N\mu} e^\beta_{_N\nu}$,
has universal symmetry  $\gamma^{_N}_{ij} \equiv {g}_{oi}{ g}_{oj}
{g}^{-1}_{oo}  - { g}_{ij}$  = $\delta _{ij}$,
that corresponds to flat 3D sub-space under arbitrary
electromagnetic and gravitational fields.

The same external electromagnetic,
$A_\mu^{\neq _N} \equiv \sum_{_K}^{_K\neq _N}  q_{_K} a_{_K\mu }$,
and gravitational,
$B_\mu^{\neq _N} \equiv -G\sum_{_K}^{_K\neq _N}  m_{_K} a_{_K\mu }$,
fields determine the proper canonical four-momentum
\begin {equation}
P_{_N\nu} \equiv m_{_N}{g}^{_N}_{\mu\nu}d{x}_{_N}^\mu/ds_{_N} =
m_{_N} \delta^\alpha_\nu V_\alpha + m_{_N}U_{\nu}^{\neq _N},
\end {equation}
where $ \delta ^\alpha _\mu V_\alpha $ = $ \{
\beta^{-1}, - \beta^{-1}v_i \}$, $\beta = \beta_{_N}$ =
${\sqrt {1 - \delta_{ij}v^iv^j}}$,
$c = 1$, $v^i = v^i_{_N} = dx_{_N}^i/d\tau_{_N} $,
$ds_{_N}^2 = d\tau_{_N}^2 - \delta_{ij}dx_{_N}^idx_{_N}^j $.

The proper time rate $d\tau_{_N}\!=\!\beta^{-1}\!ds_{_N}\!=\!{\sqrt
{g^{_N}_{oo}}}(dx^o_{_N}-g^{_N}_idx_{_N}^i)$ =
$(1+\beta U^{\neq_N}_o)d{x}_{_N}^o$  +
$\beta U^{\neq_N}_id{x}_{_N}^i = d{x}_{_N}^o +
\beta U^{\neq_N}_\mu d{x}_{_N}^\mu   =
d{x}_{_N}^o + \beta^2 U^{\neq_N}_\mu
P_{_N}^\mu m_{_N}^{-1} d\tau_{_N} $
depends on all external gravitational and electromagnetic fields,
\begin {equation}
\left ( {{d \tau_{_N}}\over dt} \right )^2   =
\left ( {  {  1 + \beta (B^{\neq _N}_o +
m^{-1}_{_N}q_{_N}A^{\neq _N}_o  ) }\over
 {1 - \beta (B^{\neq _N}_i + m^{-1}_{_N}q_{_N}A^{\neq _N}_o) v_{_N}^i } }
\right )^2.
\end {equation}

The gravitational dilation of the proper time rate
$d\tau_{_N}$ with respect to the laboratory time interval
$dt$ coincides for weak fields in (3) with the similar
result of general relativity [2].
There are also observations [4-6] of
electromagnetic time dilation-compression for charges.
Vector electrogravity explains this time relativity by
coupling gravity and electrodynamics in the "old"
pseudo-Riemannian four-space, but with electromechanical
connections, when
\begin{eqnarray}
  {g}_{oo} = (1 +  \beta U_o )^2,
    \   {g}_{oi} = (1 +   \beta U_o )
 \beta U_i, \
 { g}_{ij} = {\beta^2 }U_iU_j   + \eta_{ij},
\nonumber \\
{ g}^i = - { g}^{oi} = \gamma^{ij} { g}_j = { g}_i =
- { g}_{oi}{ g}_{oo}^{-1} =
  -U_i(\beta^{-1} + U_o )^{-1}
\nonumber \\
 g^{oo} = g^{-1}_{oo} -
 g_i g^i
= (1 - \beta^2 U_iU_j \delta^{ij})(1 + \beta U_o)^{-2}, \
\gamma_{ij} = \gamma^{ij}=  -g^{ij} = \delta_{ij}
\nonumber \\
P_\mu = m \{\beta^{-1} + U_o \ ; \ -\beta^{-1}v_i + U_i \}
   = m(\delta_\mu^\alpha V_\alpha + U_\mu) =
g_{\mu\nu}P^\nu
\nonumber \\
P^\mu =\{ m (\beta^{-1} + U^o) \ ; \ P^i  \} =  m \{\beta^{-1} -
(U_o + U_i v^i )(1 + \beta U_o)^{-1}\  ;\ \beta^{-1} v^i \}
\nonumber \\
 \ P_\mu P^\mu = g_{oo}
(P^o - g_i P^i)^2 - \delta_{ij}
 P^iP^j = P_o^2g^{-1}_{oo}
- m^2\beta^{-2} v^2    =  m^2.
\end{eqnarray}

Proper four-space $x_{_N}^\mu$, metric tensor $g_{\mu\nu}^{_N}$,
time rate $d\tau_{_N}$ and
four-interval $ds_{_N} \equiv \pm {\sqrt {
g^{_N}_{\mu\nu}dx^\mu dx^\nu}}$ may be introduced only for
one selected object N and they do not coincide with similar
proper functions of other objects. An ensemble of different
charged and neutral objects may be described in common
three-space exclusively due to universal  geometry, $\gamma^{_K}_{ij}
= \delta_{ij}$, of all their three-intervals $dl_{_K}$.
There is no universal geometry for all four-intervals, and
no one rate $d\tau_{_K}$ can be used
as a universal time interval for an ensemble of interacting
elements.  One may however employ the common time interval,
$dt \equiv \pm {\sqrt {\gamma^{_K}_{oo}dx^o_{_K}dx^o_{_K}}} =
\pm {\sqrt {\delta_{oo}dx^o_{_K}dx^o_{_K}}}$ = $\pm|dx^o|$,
which is appropriate for all matter, $dt = |dx^o|$, and antimatter,
$dt = -|dx^o|$, due to the universal metric tensor
$\gamma^{_K}_{oo} = \delta_{oo}$ of flat one-dimensional
proper intervals $|dx^o_{_K}|$. Laboratory evolution
of matter is three-dimensional because only the universal
flat intervals, $  dt_{_K}$ and $dl_{_K}$, rather than the
unique proper  four-intervals $ds_{_K}$,  have a common sense
for the total ensemble.

According to the tetrad (1) this proper four-space can
take Euclidean metric in a local inertial reference system,
where all external fields are absent or balanced,
$U_\mu^{\neq _N}(x^\nu_{_N}) = 0$.
The equivalency principle leads to the following relations,
\begin {equation}
{{D P_{_N\mu} }\over dt} = {{dx_{_N}^\nu}\over dt}
{{\partial P_{_N\mu}  }\over \partial x^\nu_{_N}},
\end {equation}
for the charged object N in its proper four-space $x^\nu_{_N}$. The
novel geodesic relations and dynamics
of charges under the canonical conservation
$P_{_N\mu}P^{\mu}_{_N} = m^2_{_N}$
in arbitrary electromagnetic fields can be tested in experiments.

A new relation between  $ds^2_{_N}\equiv {{
g^{_N}_{\mu\nu}dx_{_N}^\mu dx_{_N}^\nu}}$, $dl^2_{_N}
\equiv {{\delta_{ij}dx_{_N}^idx_{_N}^j}}$, and
$dt^2_{_N} \equiv {{\delta_{oo} dx_{_N}^odx_{_N}^o}}$
may be derived, due to (1), from the equality
$ds^2_{_N}  \equiv (dx_{_N}^o
+ \alpha_{_N} ds_{_N} )^2$ + $\delta_{ij}dx_{_N}^idx_{_N}^j$,
where $\alpha_{_N}  \equiv  \beta U_{\mu}^{\neq _N}P_{_N\mu}m^{-1}_{_N}$
$\equiv (U_o^{\neq _N} + U^{\neq _N}_i v^i )/(1 + \beta U_o^{\neq _N})$.
The pseudo-Riemannian four-interval for charged matter reads
\begin {equation}
ds^2 (\alpha_{_N}) \equiv
\left ( {{\alpha_{_N} dx^o \pm {\sqrt {(dx^o)^2 - dl^2(1-\alpha_{_N}^2)} } }
\over (1 - \alpha_{_N}^2)}   \right )^2
\approx {{dt^2}\over (1 - \alpha_{_N})^2}
- {{dl^2}\over (1-\alpha_{_N})},
\end {equation}
when $(1-\alpha_{_N}^2)dl^2/dt^2 \ll 1 $.   Notice that
there is no Schwarzschild's divergence in (6) because $\alpha_{_N} < 0$
for pure gravitational potentials.

The four-interval $s_{_N}(\alpha_{_N})$ depends on external
gravitational and electromagnetic fields that can be verified
in practice for both
neutral and charged non-relativistic objects.
Solutions of (6) with $(-\alpha_{_N}) = GM/r \ll 1$ can  explain,
for example, the measured planet perihelion precession under flat
three-space.

One may verify from (4) or (1), that the metric tensor or
its tetrad takes  four field degrees of freedom due to the
external four-potential $U_{\mu}^{\neq _N}$. Thus, the
Hilbert variation with respect to ten "independent"
components of $g^{_N}_{\mu\nu}$ is not a well-defined
procedure. There is no physical notion with ten degrees of
freedom behind the metric tensor $g^{_N}_{\mu\nu}$, which is
not an independent proper variable because
$g^{\mu\nu}_{_N} P_{_N\mu}P_{_N\nu}
= scalar$.

Due to  the  variation of the proper particle-field action $S_{_N}$,
\begin {equation}
\delta S_{_N}=-\int dx^4{\sqrt {-g}} T_{_N}^{\mu\nu}
 \delta g_{\mu\nu}(P_{_N\lambda})=-\int dx^4{\sqrt {-g}} T_{_N}^{\mu\nu}
{{\partial  g^{_N}_{\mu\nu}}\over \partial e^{\alpha}_{_N\rho} }
{{\partial  e_{_N\rho}^{\alpha}}\over \partial P_{_N\lambda}}
\delta P_{_N\lambda},
\end {equation}
gravitation has to be rewritten in terms of four-vector
contraction of the Hilbert energy tensor $T_{_N}^{\mu\nu}$.
A basic dynamic equation, which determines the forming-up four-potential
$a_{_N\mu}$, takes a vector Maxwell-type form,
$T^{\mu\nu}_{_N}P_{_N\mu}$ = 0, when $\delta S_{_N}$ = 0 in (7).
Wave solutions of this four-vector equation correspond
to photons, which are responsible for both gravitation and
electromagnetic  interactions.

By taking the proper action  $S_{_N}$ =
$-\int{\sqrt {-g}}d^4x[P_{_N\mu}i_{_N}^\mu + (f_{_N}^{\mu\nu}W^{_N}_{\mu\nu}
/16\pi)]$,
 with $i^{\mu}_{_N} \equiv \int (dx_{_N}^\mu / dp_{_N})(-g)^{-1/2}
 {\hat \delta}^4 (s_{_N})dp_{_N}$,
 $f^{_N}_{\mu\nu} \equiv \nabla_\mu a_{_N\nu} - \nabla_\nu
a_{_N\mu}$, and $W^{_N}_{\mu\nu} \equiv \nabla_\mu P_{_N\nu} - \nabla_\nu
P_{_N\mu}$, one finds the proper energy tensor
\begin {equation}
T_{_N}^{\mu\nu}
=
  {{P_{_N}^{\mu}  I_{_N}^{\nu }(x)
+ P_{_N}^{\nu}I_{_N}^{\mu }(x) }\over 2}
 + {  W_{_N\rho\lambda}(x)
\over 16 \pi}
[ g_{_N}^{\mu\nu} {f}_{_N}^{\rho\lambda} -
2g_{_N}^{\mu\rho}{f}_{_N}^{\nu\lambda}
- 2g_{_N}^{\nu\rho}{f}_{_N}^{\mu\lambda}],
\end {equation}
where the Maxwell-type four-vector
$I_{_N}^\mu \equiv i_{_N}^\mu - (4\pi)^{-1}\nabla_{\nu}f^{\mu\nu}_{_N} $
may be associated with both
kind of the extended homogeneous charges,
$m_{_N}$ and  $q_{_N}$, located on the proper
light-cone $s_{_N}(x_{_N}, \xi_{_N}) = 0 $. Variations
of the scalar action
$S_{_N}$ with respect to three proper variables,
$x^\mu_{_N}$, $P_{_N}^\mu$,
and $a_{_N\mu}$, lead to a system of vector equations, $\nabla_\mu
T^{\mu\nu}_{_N} = 0$, $T_{_N}^{\mu\nu}P_{_N\mu} = 0$,
and $\nabla_\mu W^{\mu\nu}_{_N} = 0$, respectively,
which determines dynamics of
one selected object N.
The tensor Einstein-type relation, $T_{_N}^{\mu\nu} = 0$, for
all components of the symmetric energy tensor (8) appears in
vector electrogravity only for potential (superfluid,
gauge-invariant) states of the object without energy
exchange and radiation, when $W^{_N}_{\mu\nu} = 0$ and $I_{_N}^\mu = 0$.

Measurements of the electron beam bending by static Coulomb
potential can be performed with high accuracy. These measurements
could test the dynamic equations (4)-(6)
and  the basic relation $P_{_N\mu}P_{_N}^\mu = m^2_{_N}$ for
canonical four-momentum.
Electromagnetic dilation-compression of time (3)
provides one more opportunity to test the developed
double unification (particle with field, gravity with
electromagnetism) in the laboratory.

Vector electrogravity operates only with flat three-space for both
gravity and electrodynamics.
The vector approach to gravitation
suggests the unified nature of gravitational and electromagnetic
radiation, associated with photon waves. These vector waves can
modulate only the four-space metric tensor $g^{_N}_{\mu\nu}$,
but they are irrelevant to Euclidean  metrics,
$\gamma_{ij}^{_N} \equiv \delta_{ij}$,  of laboratory
three-space under all possible experiments.

\bigskip
\noindent [1]  R. Weiss, {\it Rev. Mod. Phys.} {\bf 71}, S187 (1999).

\noindent [2] C.W. Misner, K.S. Thorne, and J.A. Wheeler,
{\it Gravitation} (San Francisco: Freeman, 1973).

\noindent [3]  C.M. Will, {\it Theory and experiment in
gravitational physics} (Cambridge: Cambridge University Press, 1993),
revised ed.

\noindent [4] E.J. Saxl, {\it Nature} {\bf 203}, 136 (1964).

\noindent [5] M.A. Tamers, {\it Nature} {\bf 339}, 588 (1989).

\noindent [6] W.A. Barker,{\it US Patent } {No 5,076,971} (1991).
\end {document}